\documentclass[pre,twocolumn,showpacs,floatfix]{revtex4}

\usepackage{graphicx}
\begin{document}

\title{Shock waves in two-dimensional granular flow: effects of rough
  walls and polydispersity}

\author{Sune H{\o}rl{\"u}ck} \affiliation{The Center for Chaos and 
  Turbulence Studies, The Niels Bohr Institute, Blegdamsvej 17, 
  DK-2100 Copenhagen {\O}, Denmark}

\author{Martin van Hecke}\altaffiliation{Present address:
  Kamerlingh Onnes Lab, Universiteit Leiden, PO Box 9504, 2300 RA
  Leiden, The Netherlands}\affiliation{The Center for Chaos and
  Turbulence Studies, The Niels Bohr Institute, Blegdamsvej 17,
  DK-2100 Copenhagen {\O}, Denmark}

\author{P. Dimon} \affiliation{The Center for Chaos and Turbulence
  Studies, The Niels Bohr Institute, Blegdamsvej 17, DK-2100
  Copenhagen {\O}, Denmark}

\date{\today}

\begin{abstract}
  We have studied the two-dimensional flow of balls in a small angle
  funnel, when either the side walls are rough or the balls are
  polydisperse.  As in earlier work on monodisperse flows in smooth
  funnels, we observe the formation of kinematic shock waves/density
  waves.  We find that for rough walls the flows are more disordered
  than for smooth walls and that shock waves generally propagate more
  slowly. For rough wall funnel flow, we show that the shock velocity
  and frequency obey simple scaling laws. These scaling laws are
  consistent with those found for smooth wall flow, but here they are
  cleaner since there are fewer packing-site effects and we study a
  wider range of parameters.  For pipe flow (parallel side walls),
  rough walls support many shock waves, while smooth walls exhibit
  fewer or no shock waves. For funnel flows of balls with varying
  sizes, we find that flows with weak polydispersity behave
  qualitatively similar to monodisperse flows. For strong
  polydispersity, scaling breaks down and the shock waves consist of
  extended areas where the funnel is blocked completely.
\end{abstract}

\pacs{45.70.Mg, 45.70.Vn}

\maketitle

\section{Introduction}  \label{intro}

\begin{figure}
  \includegraphics[width=8.5cm]{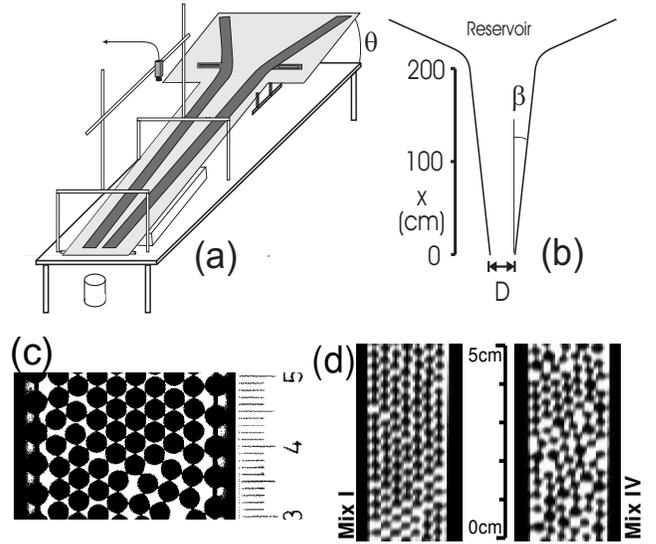}
\caption{(a) Sketch of the experimental setup.
  (b) Schematic top view of the setup, showing the important
  parameters of the funnel geometry.  (c) Close-up of the balls
  rolling between rough walls. The linked balls near the edges are
  glued to the walls (the scale is in cm). (d) Snap shots of two
  different polydisperse mixtures, referred to as mix I and IV (see
  text).}\label{f:setup}
\end{figure}

Density waves can occur in various granular flow systems such as
funnels and hoppers \cite{brown}, pipes \cite{matsushita}, and
hour glasses \cite{verveen}.  In previous experiments, we studied
the formation of kinematic shock waves which propagate against the
main flow in a two-dimensional system of balls rolling in a small
angle funnel \cite{vd96,hd99,hd00}. A sketch of the setup used is
shown in Fig.~\ref{f:setup}(a)-(b). In this earlier work, the
walls of the funnel were smooth, and the balls were of equal size
(monodisperse flow). In the present work, we study what happens
when we break the peculiarities of this smooth wall/monodisperse
system by either {\em{(i)}} making the walls of the funnel rough
(Fig.~\ref{f:setup}(c)), or {\em{(ii)}} taking ``polydisperse''
mixtures of balls of different sizes (Fig.~\ref{f:setup}(d)).

The crucial experimental parameters characterizing the flow
geometry are the funnel opening angle $\beta$ and the funnel
outlet width $D$. For smooth wall/monodisperse flows the most
important features of the shock waves were found to be the
following \cite{vd96,hd99,hd00}: {\em{(i)}} For $\beta >0$, the
rolling grains tend to locally form triangular lattices which lead
to the creation of shock waves predominantly at particular sites
in the funnel where close packing occurs.  {\em{(ii)}} The
velocities $U$ of the shocks are, in good approximation, a
function of the rescaled width $w(x)/D$ only. Here $x$ is the
coordinate along the funnel (Fig.~\ref{f:setup}(b)) and $w(x)$
denotes the funnel width at position $x$.

As we will show below, by making the walls rough or the flow
polydisperse, the triangular packing can (partially) be
suppressed. We have studied the shock statistics and the behavior
of individual balls as in previous work~\cite{hd99,hd00}, and will
present here the results for the creation and propagation of
shocks in these systems in the pipe flow (parallel walls) and in
the  small angle, intermittent flow regime~\cite{hd99,hd00}.  We
have also investigated, using ball tracking, the formation of
shear bands for rough walls and the effects of completely
stationary shock packings in polydisperse flows.

Recently, a number of related experiments on two-dimensional flow
have been performed. Tsai et al.\cite{tsai} studied a system close
to our rough wall pipe flow experiment, and included the effect of
partially blocking the outlet.  Reydellet et al.~\cite{reydellet}
studied a falling vertical column of balls where ball-ball
interactions are dominated by collisions, and the rolling of balls
does play a minor role. Finally, Le Pennec et al.~\cite{pennec}
studied two-dimensional rolling flows of small glass balls in flow
geometries with very large funnel angles.

The experimental setup and methods of analysis have been described
in detail elsewhere \cite{hd99,hd00} and we summarize the
essential aspects here. A sketch of the setup is shown in
Fig.~\ref{f:setup}(a) and the important geometric parameters are
shown in Fig.~\ref{f:setup}(b).  The balls roll on a coated Lexan
plane (with inclination $\theta=4.1^{\circ}$) in a single layer,
in a funnel formed by aluminum walls and covered by a transparent
lid.  The granular material is comprised of 50000 brass balls of
3.18~mm diameter in the monodisperse case, while details of the
polydisperse mixtures are discussed in section~\ref{poly}. The
rough walls were made by glueing linked rows of balls of nearly
the same diameter as the rolling balls to the original smooth
walls (Fig.~\ref{f:setup}(c)). The walls are straight for $200$~cm
and curve smoothly at the top to form a reservoir. They can be
moved to vary the outlet width $D$ (0-35~mm) and the funnel half
angle $\beta$ ($0^\circ-3^\circ$). A light box is placed below the
funnel to illuminate the balls from below, and a video camera is
placed above the system. Snapshots of a small part of the funnel
show the effect of rough walls (Fig.~\ref{f:setup}(c)) and
polydispersity (Fig.~\ref{f:setup}(d)) on the packing of the
balls.

The paper is structured as follows: In Section~\ref{rough} we
present data based on monodisperse flows between rough walls.
Shock wave statistics for both intermittent flows
($\beta>0^{\circ}$) and for pipe flows ($\beta=0^{\circ}$) are
discussed, and we also study data based on ball tracking.  In
Section~\ref{poly} we investigate the effect of various degrees of
polydispersity in flow between smooth walls. Finally, in
Appendix~\ref{traffic}, we discuss some similarities with traffic
flow.

\begin{figure}[b]
\includegraphics[width=8.0cm]{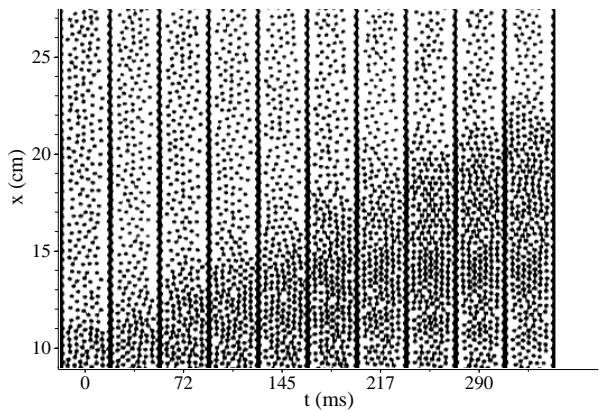}
\caption{Film sequence showing $18$~cm of the funnel showing a
propagating shock wave at $\beta=0.2^{\circ}$, $D=20$~mm (each
vertical 'stripe' is a separate picture). Such movies are recorded
at 220 frames/s and are used for the ball tracking discussed in
section~\ref{s.balltrack}. For clarity, only every eight frame is
shown.}\label{f:film}
\end{figure}

\section{Rough walls}\label{rough}

The basic phenomenology of the formation of shock waves is
illustrated in Fig.~\ref{f:film}, which shows ten subsequent
snapshots of the balls in a small section near the outlet of the
rough wall funnel. In dense regions, which may occur due to a
combination of geometric effects (the finite funnel angle $\beta$)
and the inelastic nature of the collisions, kinetic energy is
dissipated rapidly and so the balls here have a lower velocity
than the balls in dilute regions.  This leads to kinematic shock
waves, in which the balls become almost stationary and tend to
pack in a lattice which extends from wall to wall. In such a
region, both energy and momentum are efficiently transferred to
the walls.  Shocks grow in the upstream direction due to incoming,
high velocity balls, and dissolve in the downstream direction due
to the action of the (effective) gravity which eventually
accelerates the balls again. To analyze these shocks, we recorded
$\sim2-9$ second movies of $37$~cm sections of the flow (usually
at $220$ frames/s), and applied ball tracking
software~\cite{hd00}, gaining detailed knowledge of the position,
velocity and acceleration of the individual balls (see
Section~\ref{s.balltrack}).

\begin{figure}[b]
\includegraphics[width=8.5cm]{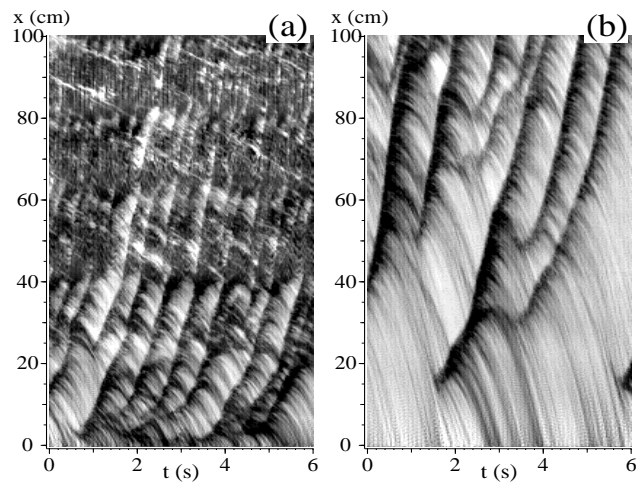}
\caption{Space-time plots of the density $\tilde{\rho}(x,t)$ at
$\beta=0.4^{\circ}$ for, (a) $D=15$~mm and (b) $D=25$~mm.  }
\label{f:Rdnsbeta}
\end{figure}

In many cases, however, one may be more interested in the overall
features of the shocks and not so much in the individual balls.
For such analysis, sequences of images taken at $60$~frames/s
and covering $100$~cm sections of the funnel were averaged in the
transverse direction to obtain a one-dimensional relative density
profile. An example of a space-time plot of this relative density
$\tilde{\rho}(x,t)$ is shown in Fig.~\ref{f:Rdnsbeta} for two
different sets of parameters.  From these the position of a shock
wave and its creation site, average local velocity $U(x)$ and
frequency $\nu(x)$ can be determined. As with smooth walls, shock
waves between rough walls are easily distinguished by
eye~\cite{hd99}.  They are created at various positions in the
funnel and there are some noticeable interactions between shocks.
As in earlier work on smooth walls, shock waves are created more
often at large funnel angles and for small values of $D$. The
shock velocity $U$ is generally observed to grow with increasing
$\beta$, $x$ and decreasing $D$.

\subsection{$\beta > 0 $}\label{Roughbeta}

We will study now in detail the statistics of shock creation,
velocity and frequency, based on density data taken for $D$
ranging from 15 to 30 mm, and for $\beta$ ranging from
$0.1^{\circ}$ up to $1.2^{\circ}$. The case of pipe flow
($\beta\!=\!0$), where rough and smooth walls show very different
behavior, will be studied in section \ref{s.pipeflow}, and some
results following from ball tracking are discussed in section
\ref{s.balltrack}.

\subsubsection{Shock creation}

The monodispersity of the balls permits them to form close packed
triangular lattices at certain packing sites $x=\chi_i$ in the
funnel. For smooth walls, the positions of these packing sites are
given by
\begin{equation}\label{eq:packsites}
\chi_i = \frac{2r + \sqrt{3} r (i-1) - D}{2 \tan \beta}~,
\end{equation}
where $i$ is an integer and $r$ is the ball radius~\cite{hd99}.  In a
funnel with smooth walls, packing effects are rather pronounced as
shown in Fig.~18 of~\cite{hd99}.

\begin{figure}[b]
\rotatebox{-90}{\includegraphics[height=8.5cm]{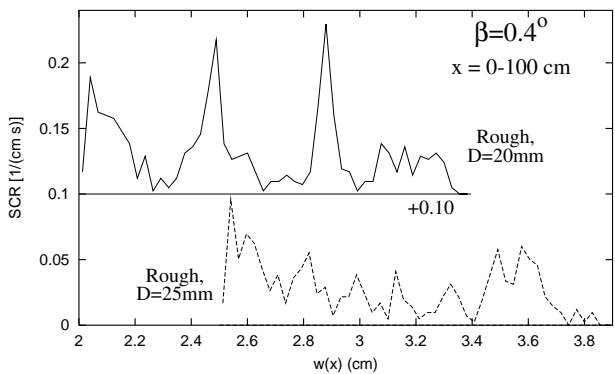}}
\caption{Manually counted shock creation rates (SCR)~\cite{SSnote}
for rough
  walls shown as a function of the local funnel width $w(x)$ for
  $D=20$~mm and $D=25$~mm (both data set cover $x=0-100$~cm).
  }\label{f:Rshockstart}
\end{figure}

For rough walls, it is slightly more difficult to give an estimate
of where one expects packing sites. It is, for example, not
obvious whether the value of $D$ (which is the minimum distance
between the two rough walls) is the relevant parameter. We have
found that packing effects persist from small $D$, but are washed
out for larger funnel outlets. This is illustrated in
Fig.~\ref{f:Rshockstart} which shows the manually counted shock
creation rates for two different values of $D$ as a function of
the local width $w$ (this aligns possible packing sites). Clearly,
packing effects are present for $D\!=\!20$~mm, but are washed out for
$D\!=\!25$~mm. This is consisted with the $D=15$~mm and $D=30$~mm data
sets, although in these cases it is difficult to obtain a precise
estimate for the shock creation rates. A comparison of the data
shown in Fig.~\ref{f:Rshockstart} to data on the shock creation
rates in smooth funnels, as shown in the top of
Fig.~\ref{f:Pshockstart}, confirms that rough walls suppress the
effects of packing.

We therefore have shown that packing effects are suppressed in
sufficiently wide rough wall flow, and that in this case shock
waves are created everywhere in the funnel with equal probability.

\subsubsection{Shock velocity }\label{RoughShockvel}

\begin{figure} [b]
\includegraphics[width=7.7cm]{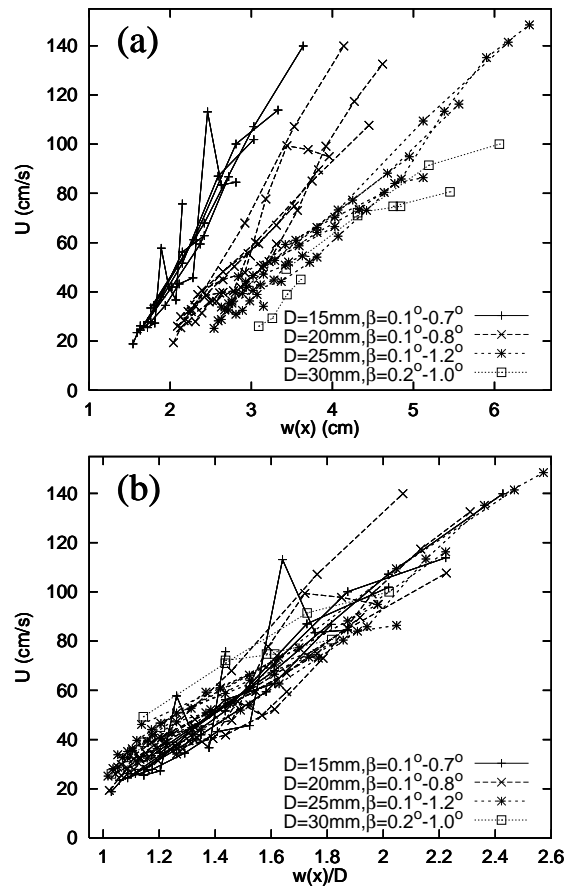}
\caption{ Average shock velocity $U(x)$
  for 40 different sets of parameters $\beta$ and $D$ as function of
  (a) the local width $w(x)$ and (b) the rescaled, non dimensionalized
  width $w(x)/D$.  }\label{f:Rshockvel}
\end{figure}

From the density fields, we can obtain the local shock velocity $U(x)$
using the shock detection algorithm RGE ({\em R}elative density
contrast {\em G}radient {\em E}dge detection) \cite{RGEnote}. This is
a refinement of the GE-method described in detail in~\cite{hd99}, and
is used to measure $U(x)$ and $\nu(x)$.  The shock velocities obtained
with this method for forty different sets of parameters are shown in
Fig.~\ref{f:Rshockvel}(a) as a function of the local funnel width
$w(x)$. This already indicates a clustering of the data in groups
given by the funnel outlet $D$.  When the data is replotted against
the dimensionless parameter $w(x)/D$ as in Fig.~\ref{f:Rshockvel}(b),
there is a fairly good data collapse with $U(x) \approx 79[w(x)/D
-0.73]$~cm/s.  A similar scaling ($U(x) \approx 64[w(x)/D-0.14]$~cm/s)
was observed for smooth walls\cite{hd99,hd00}. Experiments with one
rough and one smooth wall, which we will refer to as ``semi rough''
(not shown) gave a more disordered flow with a decent data collapse of
$U(x) \approx 83[w(x)/D-0.57]$~cm/s, which falls between rough wall
and smooth wall data for low $w(x)/D$ (see Fig.~\ref{f:linefit}(a)).  Therefore we
conclude that the roughness of the walls leads, in general, to a
slowing down of the shock waves.

In the data collapse Fig.~\ref{f:Rshockvel}(b) there seems to be a
trend for wide funnels to have slightly larger velocities.  Some
of the $D=30$~mm data sets had problems with static buildup at the
outlet during parts of the experiment which is likely to have
affected shock statistics (increasing $U(x)$ and $\nu(x)$) near
the outlet. It is also possible that for larger outlet widths the
data for rough walls becomes more comparable to those of smooth
wall systems. Nevertheless, we find that the shock velocity is in
good approximation  a function of the local rescaled width
$w(x)/D$ only, both for smooth and rough walls.

\begin{figure}
  \includegraphics[width=7.7cm]{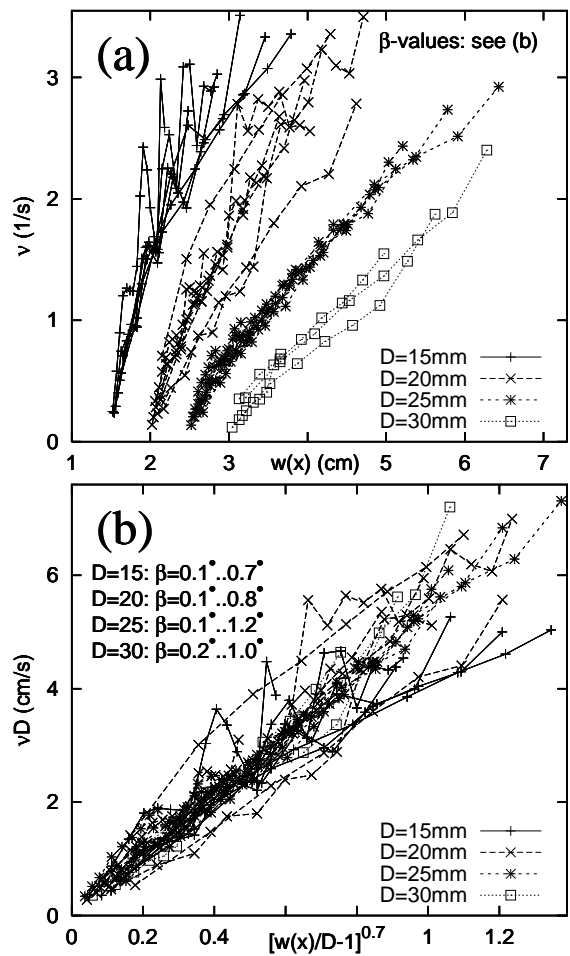}
\caption{ (a) Average shock frequency $\nu(x)$ vs. $w(x)$.  (b)
Rescaled shock frequency $\nu(x)D$ vs. $(w(x)/D)^{0.7}$ (lin/lin).
All for 40 different sets of parameters $\beta$ and
$D$.}\label{f:Rshockfreq}
\end{figure}

\subsubsection{Shock frequency}\label{RoughFreq}

From the density plots we can also obtain the local shock frequencies
$\nu(x)$ using the RGE algorithm~\cite{RGEnote}.  In
Fig.~\ref{f:Rshockfreq}(a) we show the shock frequencies for our rough
wall data as a function of $w(x)$. While some packing site periodicity
is still visible for $D=15$~mm, for larger values of $D$ the curves
are fairly smooth and we will discuss the data-collapse that occurs
there.

From the data shown in Fig.~\ref{f:Rshockfreq}(a) it appears that
$\nu$ goes to zero when $w(x) \rightarrow D$, i.e., near the outlet.
Replotting the data then as function of $w(x)/D-1$ (not shown)
confirmed this, and indicated that to achieve data collapse, one needs
to plot the product $\nu(x) D$ as a function of $w(x)/D-1$. Finally,
in contrast to the shock velocity, the frequency clearly grows
non-linearly with $w(x)/D-1$. On a log-log plot, it appears that $\nu
D$ scales as $(w(x)/D-1)^p$, with an exponent $p$ around 0.7
(Fig.~\ref{f:Rshockfreq}(b)). While the scaling range is to small to
determine whether such powerlaw scaling holds asymptotically, it is a
useful way of collapsing our data, as also is shown in
Fig.~\ref{f:Rshockfreq}(c) Note that significant deviations can be
seen for larger values of $w(x)/D$.  The data falling below the curve
in Fig.~\ref{f:Rshockfreq}(c) mainly belong to data sets with
$D=15$~mm $\beta \geq 0.5^{\circ}$ where the RGE method tends to
underestimate $\nu(x)$ for large $x$,$\beta$~\cite{RGEnote}.

In a previous paper~\cite{hd99}, which discussed the shock dynamics
for smooth funnels, the issue of $\nu(x)$ data collapse was not
discussed~\cite{Dnunote}. Reprocessing of this smooth wall data
employing the RGE algorithm does show a decent data collapse where
$\nu(x) D =1.9(1) (w(x)/D-1)^{0.7(1)}$, but the number of old data
sets with $D \neq 10$~mm that gives reliable $\nu(x)$-data is
insufficient to judge whether $\nu(x) D$ or $\nu(x)$ gives a better
data collapse. We will discuss this issue below for weakly
polydisperse mixtures in smooth funnels in Section~\ref{Polyscale},
where more data is available.

Note that the shock frequency for smooth walls is generally less than
half the frequency for rough walls. However, since we believe that
shock waves are, in general, produced by disturbances in the flow,
this should not be a big surprise: rough walls produce more
disturbances.

Data obtained from flows with one rough wall and one smooth wall
(``semi rough'') confirm this intuitive picture, since we find a
decent data collapse of $\nu(x)D \approx 3.4(1)
(w(x)/D-1)^{0.7(1)}$~cm/s (not shown). This is right between the fits
for the rough wall and smooth wall data discussed above.

\begin{figure}
\includegraphics[width=8.5cm]{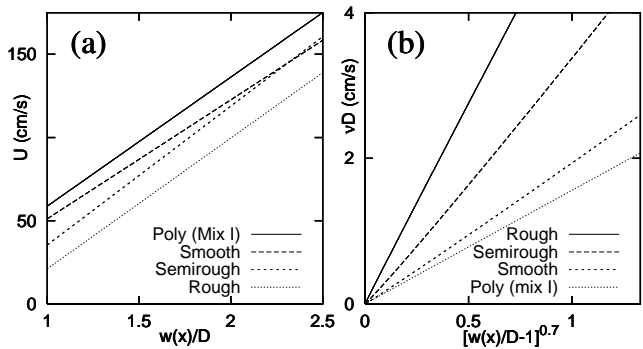}
\caption{Linear fits for $1 \leq w(x)/D \leq 2.5$ of $U(x)$ vs.
$w(x)/D$ in (a) and for $\nu(x)D$ vs. $w(x)/D$ in (b). Both graphs
show the lines for smooth wall, smooth/rough combined walls and
rough wall flows. Data for weakly polydisperse, smooth wall flows
are also included.}\label{f:linefit}
\end{figure}

\subsubsection{Rough wall intermittent flow: conclusion}

The rough walls allow us to suppress packing effects, so that two
simple scaling laws for the shock velocity and frequency emerge:
\begin{eqnarray}
U(x) &\approx& \alpha_1 (w(x)/D- \alpha_2) \label{sc1}\\ \nu(x)
&\approx& D^{-1} \alpha_3 (w(x)/D-1)^{0.7} \label{sc2}
\end{eqnarray}
The fitting coefficients $\alpha_1,\alpha_2$ and $\alpha_3$ vary with
the roughness of the walls. In Fig.~\ref{f:linefit} we show the best
fits for data obtained for rough walls, semi-rough walls, smooth walls
and smooth walls / weakly polydisperse balls (discussed in detail in
Section~\ref{Polyscale}). The fits are all made in the interval $1
\leq w(x)/D \leq 2.5$ for consistency.

As shown in Fig.~\ref{f:linefit}(a), changes in the boundaries have a
weak effect on $\alpha_1$, but substantially affect $\alpha_2$.  The
old smooth wall data for the velocity seems to deviate a little from
this trend, and we do not know the reason for this.  For the
frequency, the roughness of the walls has a profound influence on the
prefactor $ \alpha_3$. For small angle funnels, rough walls promote
the occurrence of shock waves, but do not affect their propagation
velocity substantially.

\subsection{Pipe flow}\label{s.pipeflow}

The behavior of shocks in pipe flow ($\beta=0^\circ$) can be expected
to be rather different from those in intermittent flow ($\beta \geq
0.1^{\circ}$) since for pipe flow there is no ``geometric'' source for
the formation of shock waves, and dissipative collisions are now
presumably the dominant origin of shock wave formation.

Indeed, for smooth walls the (few) shock waves that are generated
never reach the reservoir and consequently the flow rate (and
indirectly all other flow properties) are entirely determined by the
reservoir outflow 
%- thus making the pipe flow qualitatively different
%from flows with $\beta \geq 0.1^{\circ}$ 
\cite{vd96,hd99}. Surprisingly, the qualitative features of pipeflow
with rough walls are similar to the intermittent flows. For example,
many shocks do travel upwards and reach the reservoir. 
%For smaller
%outlet widths, shock waves appear almost everywhere, but do not
%persist for more than $\sim 20$~cm or $\sim 5$~secs, while for larger
%outlet widths there are far fewer shock waves (in pipe flow the shock
%frequency $\nu$ is generally observed to drop with growing $D$), but
%they persist for at least $\sim 100$~cm.

%\begin{figure}[b]
%\includegraphics[width=8.5cm]{fig08_R_dnspipe.ps}
%\caption{Space-time diagrams for the density $\tilde{\rho}(x,t)$
%for pipe flow ($\beta=0.0^{\circ}$) in the upper half of the
%funnel ($x=100-200$~cm) shown for (a) $D=15$~mm and (b) $D=30$~mm.
%(Please note compressed x-axis relative to
%Fig.~\ref{f:Rdnsbeta}.)}\label{f:Rdnspipe}
%\end{figure}

% Two examples of the relative density for a pipe
%flow with rough walls are shown in Fig.~\ref{f:Rdnspipe}. Here,
%the qualitative features are similar to the intermittent flows.
%For example, many shocks do travel upwards and reach the
%reservoir. For the smaller outlet width in
%Fig.~\ref{f:Rdnspipe}(a), we see that shock waves appear almost
%everywhere, but do not persist for more than $\sim 20$~cm or $\sim
%5$~secs.  Their velocities are of the order of $U\sim 15$~cm/s.
%For the larger outlet width in Fig.~\ref{f:Rdnspipe}(b), there are
%far fewer shock waves (in pipe flow the shock frequency $\nu$ is
%generally observed to drop with growing $D$), but they persist for
%at least $\sim 100$~cm, and  their velocities are very similar to
%those of small channel widths.
 
Similar to smooth wall pipe flow, the rough wall pipe flow is
extremely sensitive to exact experimental conditions and consequently
the statistics obtained for these flows is bound to be more noisy. In
fact, as we will see below, subtle problems near the outflow area
render part of the data unusable. The analysis of shock properties is
further hampered by the fact that the RGE shock identification
method~\cite{RGEnote} used above has problems detecting some shocks
with $U<15-20$~cm/s.  We have therefore decided to use an alternative
method, based on space-time correlation functions, to obtain a measure
of the velocity, and determine their frequency based on a simple
threshold algorithm (see Appendix B).

\begin{figure}
\includegraphics[width=8.0cm]{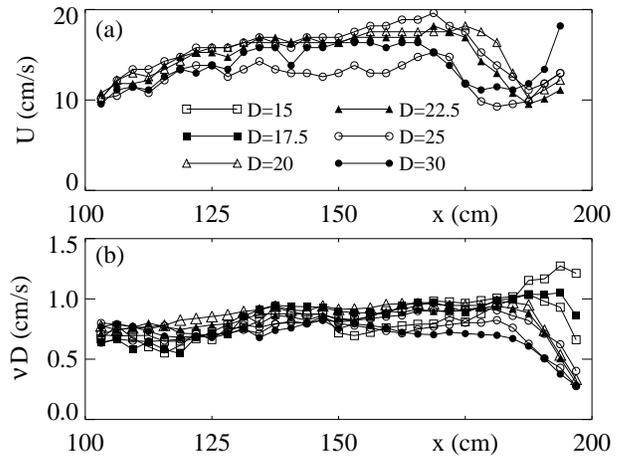}
\caption{Velocity (a) and rescaled frequency (b) of shock waves in
rough wall pipe flow.}\label{f:pipe_results}
\end{figure}

Based on these methods, we have determined the shock velocities for
rough wall pipe flow. We find that the data for the lower part of the
funnel ($0<x<100$~cm) does not show any systematic trend, and often
here data sets show a complicated mixture of periods of stationary
shocks and moving shocks. We think that this is due to experimental
problems (in particular static charge buildup near the funnel outlet
in some measurements).  In later runs, when we studied the upper
funnel ($100<x<200$ cm), these problems were solved. Here we find that
the shock velocity is essentially independent of the width $D$, apart
from some effects near the outflow of the reservoir (see
Fig.~\ref{f:pipe_results}(a)).

The shock frequencies decrease with $D$, but do not show a clear trend
with $x$. Rescaling the frequency with $D$, we find that $\nu D$ is
fairly constant, for $D$ ranging from 15~mm to 30~mm (see
Fig.~\ref{f:pipe_results}(b)). Again, most data for ($0<x<100~ cm$)
does not show any systematic behavior, due to the large amount of
stationary shocks.

Using the linear fits of $U(x)$ and $\nu(x)D$ sinilar to
Section~\ref{Roughbeta}, we find $U(w(x)=D) \approx 21$~cm/s and
$\nu(w(x)=D)D \approx 0.8$~cm/s.  Considering the systematic
differences between RGE based and space/time correlation based
measurements of $U(x)$ there is a fairly good correspondence between
the pipe flow shock wave data from the $x>100$~cm part of the funnel
and the shock wave behavior for $\beta>0^{\circ}$ flows. This suggest
that the mechanisms of shock wave propagation are not too different
after all.
%, but that in the pipe flows it represent a
%kind of sensitive equilibrium that is hard to establish near the
%outlet.

\begin{figure}
\includegraphics[width=8.5cm]{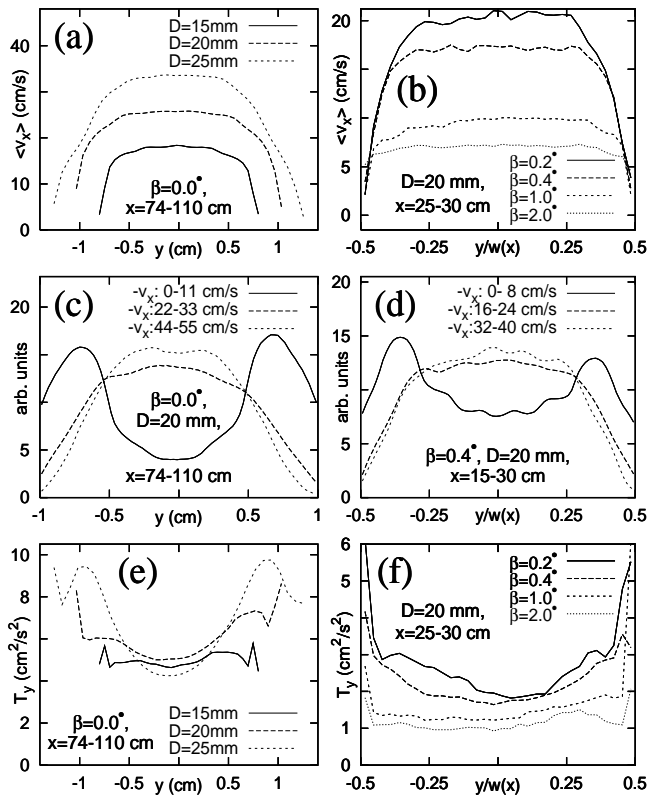}
\caption{Average $v_x$ as function of $y$ in various pipe flows
(a) and $\beta \geq 0.1^{\circ}$-flows (b). Histograms of ball
positions (considering the width of the balls) grouped according
to the $v_x$-value for a pipe flow ($\beta=0.0^{\circ}$,
$D=20$~mm) in (c) and intermittent flow ($\beta=0.4^{\circ}$,
$D=20$~mm) in (d). The average square transverse velocity $<v_y^2>
= T_y$ is shown as function of $y$ for three pipe flows in (e) and
for $\beta
> 0^{\circ}$ flows in (f). Note that in (b),(d) and (f) the rescaled coordinate
$y/w(x)$ is used on the horizontal axis. }\label{f:Ryprof}
\end{figure}

We conclude that in good approximation, rough wall shock flow is
dominated by shocks that are created mainly near the outflow region,
travel upward with fairly constant velocity and whose frequency
decreases as $1/D$.

\subsection{Ball tracking}\label{s.balltrack}

%\begin{figure}
%\includegraphics[width=8.5cm]{fig11_R_acccomp.ps}
%\caption{Comparison of the 1D-acceleration field $a(x,t)$ of (a) a
%smooth wall flow at $\beta=1.0^{\circ}$, $D=14$~mm,
%(b) a rough wall flow at $\beta=1.0^{\circ}$, $D=15$~mm .
%}\label{f:Racccomp}
%\end{figure}

We have processed detailed films such as shown in Fig.~\ref{f:film} to
obtain ball trajectories (details of this method are described in
\cite{hd00}).  From the full set of ball trajectories, we have
constructed continuous one-dimensional Eulerian fields of the relative
density $\tilde{\rho}(x,t)$, velocity $v(x,t)$, and acceleration
$a(x,t)$\cite{hd00}. A comparison of the acceleration fields for rough
and smooth walls indicates that the flow for smooth walls has more
disturbances and the shocks are less sharply defined.
% The
%acceleration field for configurations with rough and smooth walls
%as shown in Fig.~\ref{f:Racccomp} illustrates that the flow for
%smooth walls has more disturbances and the shocks are less sharply
%defined.

We have previously shown (Appendix C in~\cite{hd00}) that smooth
wall flows are reasonably one-dimensional, i.e., the density,
velocity and acceleration fields do not show a strong dependence
on $y$, the coordinate across the funnel. This is no longer true
for rough wall flows as shown in Fig.~\ref{f:Ryprof}. The velocity
$<v_x>$ as a function of the transverse coordinate $y$ is shown
for various pipe flows in Fig.~\ref{f:Ryprof}(a) and for $\beta >
0^{\circ}$ flows in Fig.~\ref{f:Ryprof}(b). In both cases do we
find that $<v_x>$ drops off near the boundaries, and this effect
is most pronounced for pipe flow and small angle flow. A number of
different flow behavior could underly these statistics, the most
obvious being: (i) The flow has a shear component near the walls,
i.e., balls near walls move typically slower than in the bulk.
(ii) In shock packings the flow reaches from wall to wall, but in
fast regions (between shocks) balls near walls are repulsed.
%This
%tendency can be seen in Fig.~\ref{f:film}.
In this interpretation, balls in fast regions are repelled from
the boundaries, and this transverse momentum is absorbed via
ball-ball interactions in the funnel center, leaving the regions
near the walls relatively empty.

To resolve this ambiguity we plot histograms of the $y$ positions
of balls with their $v_x$ in a certain interval in
Fig.~\ref{f:Ryprof}(c) (for pipe flow) and in
Fig.~\ref{f:Ryprof}(d) (for $\beta = 0.4^{\circ}$ flow). If reason
(ii) would dominate, the histogram of the slowest balls (in shock
regions) should be fairly flat, but this is not the case and
therefore we conclude that there is an important shear component
to the flow.

The $y$ dependence of the $y$ component of the granular
temperature, $T_y := <v_y^2>$, is shown in Fig.~\ref{f:Ryprof}(e)
for pipe flow and in Fig.~\ref{f:Ryprof}(f) for $\beta >
0^{\circ}$ flow. For the $D=15$ mm pipe flow, $T_y(y)$ is constant
while the wider pipe flows exhibit a more quiet region in the
center -- perhaps consistent with the ``transverse momentum sink''
mentioned in (ii) above. The $\beta > 0^{\circ}$ flows in
Fig.~\ref{f:Ryprof}(f) are denser and slower, showing that in
dense flows ($\beta = 1^{\circ}$, $\beta = 2^{\circ}$) transverse
momentum is absorbed immediately and the profile is flat. The
``transverse heat sink'' does not play a role here.

\section{Polydisperse flows}\label{poly}
\begin{table}[b]
\begin{ruledtabular}
\begin{tabular}{llllll}
Mixture & 2.5 mm & 3.0 mm & 3.2 mm & 3.5 mm & 4.0 mm\\
\hline
I    &       & 10000 & 10000 &      &\\
II   &       & 10000 & 10000 & 5000 &\\
III  & 10000 & 10000 & 10000 &      &\\
IV   &       & 10000 & 10000 &      & 3000\\
\end{tabular}
\end{ruledtabular}
\caption{\label{tab:table1} Quantities of balls of various sizes
in the four different polydisperse mixtures used here. }
\end{table}

To suppress the close packing effects that occur for monodisperse
balls rolling in smooth funnels, we have also explored flows of balls
of mixed sizes. We have studied shock waves for four different
mixtures, which we will refer to as mixture I-IV in increasing order
of ``polydispersity'' (see table 1).

\begin{figure}[t]
\includegraphics[width=8.5cm]{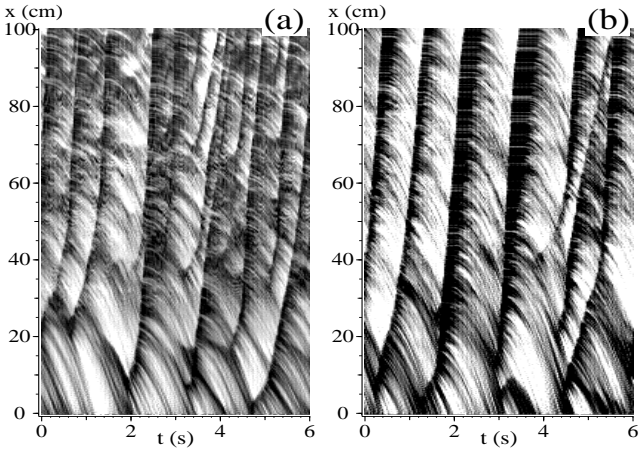}
\caption{Space-time diagrams of the density $\tilde{\rho}(x,t)$
  showing differences in shock behavior between weak and strong
  polydispersity at $\beta=0.5^{\circ}$, $D=10$~mm for Mix I (a) and
  Mix IV (b).}\label{f:Pdns}
\end{figure}

A crucial and unexpected effect of polydispersity can be seen in
the space-time plots of the density such as Fig.~\ref{f:Pdns}.
For a weakly polydisperse mixture (mix I in Fig.~\ref{f:Pdns}(a))
the shocks appear to display similar behavior as for monodisperse
balls, but for increasingly polydisperse mixtures, shock waves
lead to the {\em
  blocking of finite fractions} of the funnel.  Here all balls are
stationary for a finite time interval. Some clear examples of this
``freezing'' can be see for mix IV in Fig.~\ref{f:Pdns}(b), for $x
\approx 80~cm$ and $t\approx 3.5~s$.  The occurrence of finite blocked
fractions can be interpreted in terms of a competition between the
leading and trailing edge of a shock.  For strongly polydisperse
mixtures, the velocity of the trailing edge of the shock where the
shocks dissolve, is lower than that of the leading edge where the
shock grows due to incoming balls. As a function of its lifetime,
a shock will therefore spread out, and finite regions of the
funnel will be blocked. In contrast, when one would generate an
extended blocked area in a monodisperse flows, it appears that the
velocity of the trailing edge is larger than that of the leading
edge. Such a shock would then shrink, until leading and trailing
edge come very close together and the shock looses its spatial
extent.

\begin{figure}
  \includegraphics[width=8.5cm]{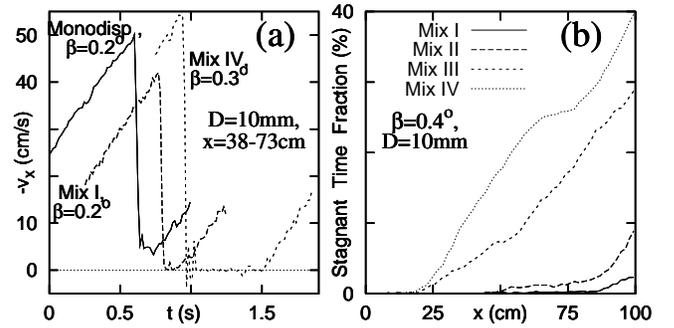}
\caption{(a) The $-v_x$(t) of individual balls during the passing of a
shock. (b) Fraction of time the flow stands still behind a just passed
shock, based on density map data
 for polydisperse Mixtures I,II,III,IV.\\
}\label{f:Pstag}
\end{figure}

We have measured, employing ball tracking methods, the velocities
of individual balls in shocks, and some representative results are
shown in Fig.~\ref{f:Pstag}(a). The ball tracking method has not
been specifically fine tuned for polydisperse flow, which leads to
a slightly noisier determination of $v_x$. The overall picture
that emerges is that in monodisperse flows, balls almost never
stop completely in a shock (notice that for the example in
Fig.~\ref{f:Pstag}(a) the minimum velocity of the balls in
monodisperse shocks indeed stays finite). In Mix I, complete
stopping of balls in a shock occurs only in a minority of the
shocks, but its occurrence increases with polydispersity and $x$:
In Mix III complete stopping of balls occurs in most shocks that
have propagated for more than $20-40$~cm and do not have another
shock right in front of them and in Mix IV blocking occurs
essentially in all shocks.

To quantify this behavior further, we have measured (based on density
space-time diagrams), for fixed values of $\beta$ and $D$ the fraction
of time that the balls are stuck in such a stationary shock as a
function of $x$ (see Fig.~\ref{f:Pstag}(b)). This data confirms what
we already observed in Fig.~\ref{f:Pdns}: the amount of blockage
increases both with $x$ and with the strength of the polydispersity.
The increase with $x$ can be understood simply from the observation
that for strongly polydisperse mixtures shocks spread out during their
lifespan, and since shocks travel upwards, the amount of blocked
channel grows with $x$.

One possible explanation we can find for the increase of this
blocking with polydispersity is the occurrence of 3D effects. One
can imagine that in a shock wave, bigger balls that are squeezed
between small balls are lifted from the support on which they
roll. When such shock dissolves, small balls have to move over a
{\em finite} distance before the bigger balls can start to roll,
leading to a finite blocking time.

\begin{figure}
\rotatebox{-90}{\includegraphics[height=8.5cm]{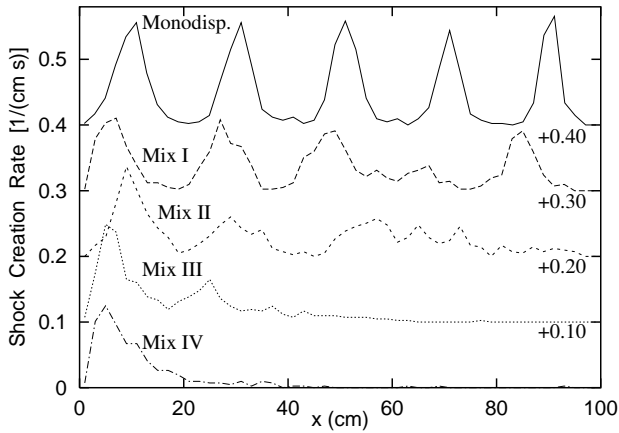}}
\caption{Shock creation rates for monodisperse flow and polydisperse
mix I, II, III, IV. for $\beta=0.4^{\circ}$, $D=10$~mm.}\label{f:Pshockstart}
\end{figure}

\subsubsection{Shock creation}

The polydispersity has two effects on the shock creation. One could
have anticipated that for stronger polydispersity the periodic packing
sites become irrelevant, but as we see from Fig.~\ref{f:Pshockstart},
what happens in addition is that all shocks are generated near the
outflow of the funnel, an effect that can also be observed when
comparing the density space-time diagram Fig.~\ref{f:Pdns}.

\subsubsection{Breakdown of scaling of $U(x)$ and $\nu(x)D$}\label{Polyscale}

\begin{figure}[t]
\rotatebox{-90}{\includegraphics[height=8.5cm]{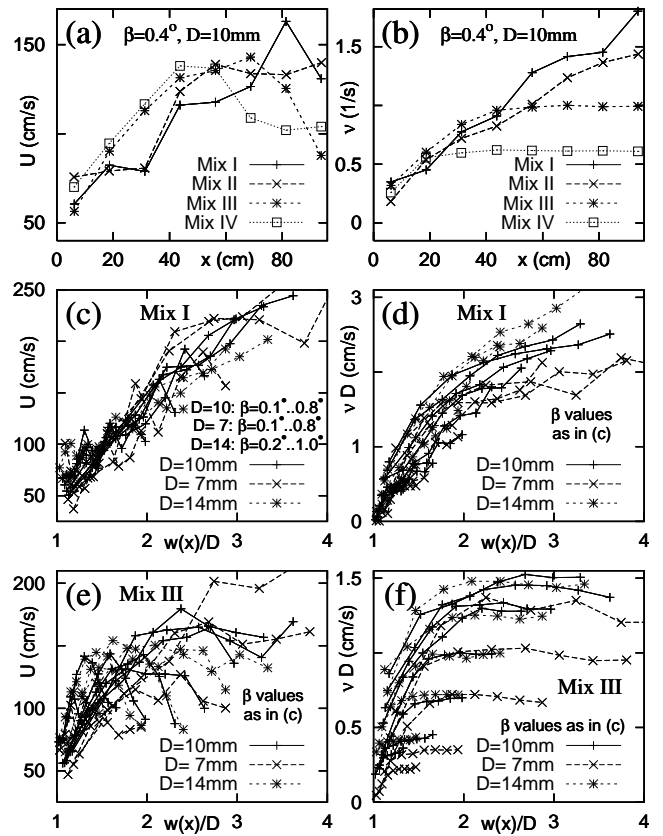}}
\caption{RGE based plots of $U(x)$ (a,c,e) and $\nu(x)$ (b,d,f):
(a) shows $U(x)$ vs. $x$ at $\beta=0.4^{\circ}$,  $D=10$~mm for
Mix I-IV. (b) shows $\nu(x)$ vs. $x$ at
$\beta=0.4^{\circ}$,$D=10$~mm for Mix I-IV. (c) shows $U(x)$ vs.
$w(x)/D$ for Mix I. (d) shows $\nu(x)D$ vs. $w(x)/D$ for Mix I.
(e) shows $U(x)$ vs. $w(x)/D$ for Mix III. (f) shows $\nu(x)D$
vs. $w(x)/D$ for Mix III. (In each of (c-f) 18 data sets are
displayed.)}\label{f:Pscaling}
\end{figure}

Using the RGE method we have studied to what extend the scaling of
$U(x)$ and $\nu(x)D$ with $w(x)/D$ holds.  In
Fig.~\ref{f:Pscaling}(a)-(b) we have plotted $U(x)$ and $\nu(x)$
for $\beta \!=\!0.4^{\circ}$ and $D=10$~mm for the four
polydisperse mixtures. For mix I, both $U(x)$ and $\nu(x)$ grow
fairly linearly with the usual packing site related
variations~\cite{hd99}. For mix II the behavior is similar but
both $U(x)$ and $\nu(x)$ become flatter for large $x$.  For the
stronger polydisperse mixtures III and IV strong deviations from
the monodisperse or weakly polydisperse case are observed. The
shock velocity $U(x)$ grows rapidly for small $x$, then peaks and
subsequently begins to drop for larger $x$.  The shock frequencies
$\nu(x)$ in mix III and IV behave similar to that of
monodisperse/weakly polydisperse flows at small $x$ but become
constant at larger $x$. This is fully consistent with the data in
Fig.~\ref{f:Pshockstart} that shows that for strong polydisperse
mixtures, all shocks are created near the outlet.

Based on the similarity between weak polydisperse and monodisperse
flows we plot $U(x)$ vs. $w(x)/D$ for mix I in
Fig.~\ref{f:Pscaling}(c). Apart from packing site variations the
data collapse is reasonably good, so in this respect the weakly
polydisperse flow behaves as a monodisperse flow.  Linear fits of
the data in Fig.~\ref{f:Pscaling}(c) yield $U(x) \approx
74[w(x)/D-0.18]$~cm/s.  It is perhaps surprising that the mix I
shock velocities are higher than those found for smooth wall
monodisperse flows (see Section~\ref{Roughbeta}).

The frequency deserves some more attention.  When we plot
$\nu(x)D$ vs. $w(x)/D$ in Fig.~\ref{f:Pscaling}(d) the data
collapse is not very convincing, especially for large $x$ where
the rescaled frequency is widely spread.  This may be due to
beginning effects of polydispersity (causing $\nu$ to drop for
high $x$ as shown in Fig.~\ref{f:Pscaling}(b)) or partial failure
of the RGE algorithm. This algorithm tends to discard weak shocks
at high $\beta$, $x$ since the density contrast of the shocks
becomes very small there (See~\cite{hd99,RGEnote}).  Since
$D=7$~mm data (highest shock frequency and lowest density
contrast) show the biggest deviation the latter reason may be the
most important.

Despite the deviations for high $x$ there seems to be a data
collapse of $\nu(x)D$ vs. $w(x)/D$ which is superior to $\nu(x)$
vs. $w(x)/D$: Together with the re-analysis of the older data
\cite{hd99} discussed in Section~\ref{RoughFreq} this lead us to
believe that the scaling of $\nu(x)D$ with $w(x)/D$ is a feature
of both rough wall and smooth wall flows. Power Law fits of the 
data in Fig.~\ref{f:Pscaling}(d) yield $\nu(x)D \approx 
1.6(1)(w(x)/D-1)^{0.7(1)}$~cm/s. That the shock frequencies of mix I are
slightly lower than for monodisperse flows is not surprising,
since polydispersity generally seems to make it harder for shock
waves to form anywhere else than near the outflow of the funnel.

For the $U(x)$, $\nu(x)$ statistics based on mix III density data
shown in Fig.~\ref{f:Pscaling}(e)-(f) there is data collapse for
neither $U(x)$ nor $\nu(x)D$ vs. $w(x)/D$.  All $U(x)$ curves in
Fig.~\ref{f:Pscaling}(e) show the same pattern of rapid growth
towards a maximum value after which a moderate decline sets in. We
have found no clear pattern in the $x$  and $U$ values of the
peaks, and there seems to be no $w(x)/D$ scaling involved.  The
$\nu(x)D$ data for mix III shown in Fig.~\ref{f:Pscaling}(f)
displays growth at low $x$ followed by a plateau, but there is no
clear trend in this plateau value.  It seems likely that both
types of deviation from the monodisperse scaling relations are
linked to the stagnant regions behind shocks and thus to the
3-dimensional packing effects discussed above.

The $U(x)$ and $\nu(x)$ data for polydisperse mix II (not shown)
are similar more to mix I, while   the mix IV data (not shown) are
similar to the mix III data.

In conclusion, we find that for sufficiently strong
polydispersity, the nature of the shock waves changes
qualitatively, and that scaling relations that hold for
monodisperse flows in either smooth or rough funnels break down.

\section{Discussion}

This work, in combination with earlier work on smooth wall/
monodisperse flows \cite{hd99,hd00}, leads to a number of conclusions
about the effects of funnel geometry, wall roughness and inelastic
dissipation. It is well-known that the dissipation occurring in
ball-ball and ball-wall collisions is enhanced by the rolling nature
of the ball motion \cite{rolling}; however, the precise value of the
effective coefficient of restitution is presumably not of large
importance for the phenomenology.  In particular, the frequency and
velocity scaling for monodisperse and weakly polydisperse systems are
very similar (see Fig.~\ref{f:linefit}), even though for weakly
polydisperse flows dissipation seems enhanced. For strongly
polydisperse flow, where dissipations seems very strong, we
unfortunately loose the 2D nature of the experiment.

In the case of smooth walls and monodisperse flows, packing effects
become very important, and they tend to obscure scaling law.  Possibly
the simplest case is then the combination of rough walls and
monodisperse flows (or, to a lesser degree, smooth walls and weakly
polydisperse flows). The rough walls have the additional advantage
that they make the system less sensitive to small perturbations (which
are inevitably present for ``smooth'' walls).

The scaling laws for shock velocity and frequency,
Eqs.~\ref{sc1}-\ref{sc2}, are the main result of our work. Since these
laws also approximately hold for smooth walls etc, their main origin
must lie in the geometry of the experiment. That the shock velocity
and frequency depend on $x$ via $w(x)/D$ is not surprising, since this
is the most obvious way in which $x$ can be made non-dimensional (we
do not expect the ball diameter to play an important role). The fact
that $\nu(x) D$ and not $\nu$ scales is harder to understand, and may
point at certain relevant velocity scales (note that both $U$ and $\nu
D$ have the dimension of velocity). Two dimensional quantities that
characterize the system are the effective ball-acceleration
$a_{\mbox{eff}}$ (related to the inclination) and $D$, but these two
quantities alone are not sufficient to provide for the correct scaling
factors, since a velocity scale would be $\sqrt{D~ a_{\mbox{eff}}}$,
and a frequency scale would be $\sqrt{ a_{\mbox{eff}}/D}$, which both
show scaling with $D$ different from what is observed.

We believe that the underlying reason for the occurrence of the
scaling laws is an important open question that deserves further
study. Our data indicates that details of the ball-ball or ball-wall
interactions are not important (although rough walls lead to more
frequent shocks, they do not alter the nature of the scaling), which
suggests that a relatively straight-forward model may capture the
phenomenology here. Indeed, for pipeflow, where the geometrical cause
of shock formation dissapears, leads to quite fragile behavior, which
only in the case of rough walls seems to reproducable and similar to
small funnel angle behavior.
 
However, to make such a model one seems to need additional information
about the relation between ball densities and velocities on the one
hand, and shock frequencies and velocities on the other.

Finally, shock waves have also been studied in some other
geometries. Tsai et al.~\cite{tsai} examined a two-dimensional rolling
pipe flow of $3.2$~mm steel spheres between rough walls. Partial
blockage of the funnel outlet was used as an additional system
parameter.  These authors performed high resolution ball tracking
measurements of small ($12$~cm) sections of the funnel and obtained
detailed local measurements of $v(x,t)$, $\tilde{\rho}(x,t)$,
$T_y(x,t)$, and $v_x(y)$. For an almost unblocked flow with $D=19$~mm
they found $\nu \approx 0.4$~1/s and $U \approx 14$~cm/s, which is
consistent with our data.

Reydellet et al.~\cite{reydellet} studied $1.5$~mm metallic balls in a
vertical pipe flow with rough walls. The balls were not rolling on a
support, and the rolling of balls supposedly played less of a role
than in our experiments. They qualitatively observed the existence of
upwards propagating shock waves, using a ``double flash technique'' to
study local ball velocities and made measurements similar to those in
our Fig.~\ref{f:Ryprof}. Note that their funnel was not continuously
refilled from a reservoir and thus their studies were limited to the
transient behavior. All our experiments are preceded by a
$>30$~sec. flow in order to avoid such transient behaviors.

Finally, Le Pennec et al.~\cite{pennec} studied a 2-dimensional
rolling flow of 1~mm glass balls in flow geometries with very large
$\beta$ (mostly $30^{\circ}$), large $D$ (typically 10-120 ball
diameters) at various flow plane inclinations. Due to the different
geometry it is hard to make any direct comparison, although these
authors did measure shock velocities and frequencies.

\subsection*{Summary}

The packing effects that we observed in earlier work on monodisperse,
smooth walled flow \cite{hd99,hd00} can be suppressed by either making
the sidewalls rough or using polydisperse mixtures of balls.  For
rough walls, we find that there are more shock waves than for smooth
walls as a result of the greater disturbances in this flow. In earlier
work~\cite{hd99} on smooth wall flows we found that the average shock
speed $U(x)$ scales with $w(x)/D$.  We have found a similar scaling
for rough walls and observed that also $\nu(x)D$ scales with $w(x)/D$
for both rough and smooth walls \cite{Dnunote}.  By using ball
tracking methods we shed light on the shear flow properties for rough
walled flows.

In polydisperse flows between smooth walls we find that our scaling
relations persist for very weak polydispersity, but breaks down for
stronger polydispersity. This breakdown is most likely caused by weak
3-dimensional effects in the packing of shock waves which apparently
lead to partial blocking of the funnel. By using ball tracking methods
we shed light on the shock structure and on the extended stationary
shock packings found in strong polydisperse flows.

\acknowledgments

The authors wish to thank Christian Veje for making the original rough
walls. P.D. would like to thank Statens Naturvidenskabelige
Forskningsr{\aa}d (Danish Research Council) for support.  MvH would
like to thank CATS for continuing hospitality.

\appendix*

\section{Comparison with traffic flow}\label{traffic}

To compare our flows to traffic flows, small angle, or preferably
pipe flows should be considered. In earlier work we made
comparisons between smooth wall funnel flows with nonzero $\beta$
and traffic flows~\cite{hd00}. In smooth wall pipe flows, shock
waves are extremely fragile, and such a comparison is not
meaningful. For the rough wall pipe flow discussed here, however,
such a comparison can be made in principle. In addition, we will
discuss our results for rough wall funnel flows in terms of
traffic flows.

It is commonly assumed that there are three flow types in traffic
flow, namely, {\em uncongested flow} (a steady flow of vehicles at
low to moderate densities, {\em queue flow} (a slow flow of near
maximum density), and {\em queue discharge} (a flow of vehicles
accelerating out of a queue flow) \cite{wolf}. Traffic data are
usually represented as $v(q)$ (speed/flow relation) or as
$q(\rho)$ (fundamental diagram), where $q=\rho v$ is the flow
rate.

\begin{figure}[t]
\includegraphics[width=8.5cm]{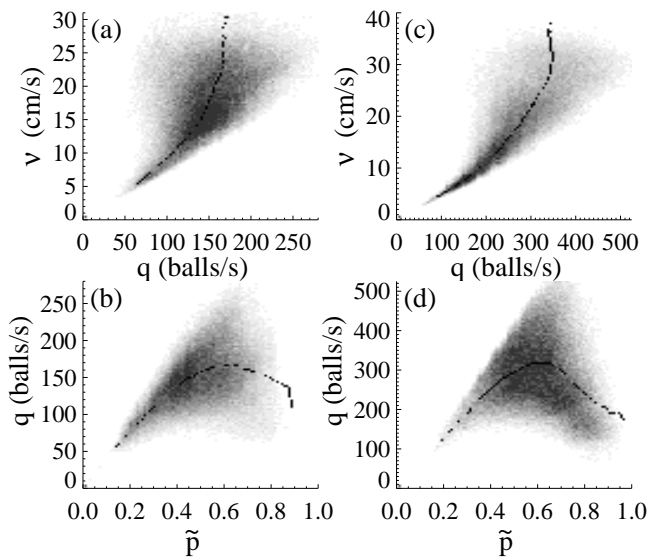}
\caption{Fundamental diagrams in rough wall flows. (a-b) Pipe flow
studied in the interval $37<x<74$ cm for $\beta=0.0^{\circ}$ and
$D=15$~mm. (c-d) Intermittent flow studied in the interval $0<x<37$ cm for
$\beta=0.4^{\circ}$ and $D=20$~mm.}\label{f:traffic}
\end{figure}

In Fig.~\ref{f:traffic}(a)-(b) we show grey scale histograms of
$(q,v)$ and $(\tilde{\rho},q)$ for a rough wall pipe flow
($\beta=0^\circ$, $D=15$~mm). Note that we use the relative
density $\tilde{\rho}$ instead of $\rho$.  The average values,
corresponding to $v(q)$ and $q(\tilde{\rho})$, are shown as solid
lines. In Fig.~\ref{f:traffic}(a), we observe regions
corresponding to queue flow and queue discharge, but there is no
region corresponding to uncongested flow. As shown in
Fig.~\ref{f:traffic}(b), the flow rate $q(\tilde{\rho})$ has a
parabolic shape with a peak value around $\tilde{\rho}=0.6$.

For larger values of $D$ (not shown), the flow rates exhibit similar
parabolic behavior, albeit with larger typical values of $v$ and $q$,
and with the $q(\tilde{\rho})$ ``parabola'' skewed towards lower
$\tilde{\rho}$ ($\tilde{\rho}=0.4-0.5$ for $D=20$~mm,
$\tilde{\rho}=0.2-0.4$ for $D=25$~mm).

Figs.~\ref{f:traffic}c-d show the corresponding diagrams for small
angle, rough wall funnel flow at $\beta=0.4^{\circ}, D=20$~mm. We
find queue flow and queue discharge but no uncongested flow,
similar to what we found in earlier work on smooth wall funnel
flow \cite{hd00}. In Fig.~\ref{f:traffic} we observe a parabolic
shaped $q(\tilde{\rho})$. Similar plots for higher $\beta$ (not
shown) exhibit ``parabolas'' skewed towards higher values of
$\tilde{\rho}$.

\section{Velocity and frequency determination for pipeflow}
\begin{figure}[t]
\includegraphics[width=8.5cm]{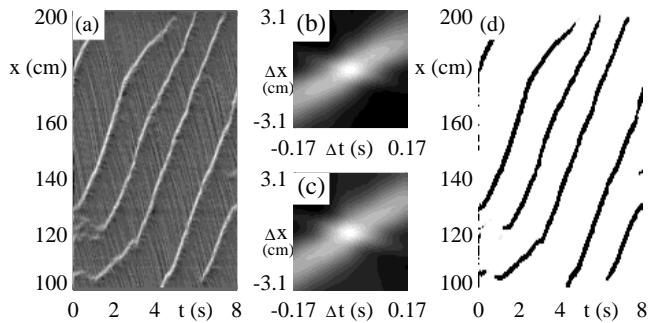}
\caption{Illustration of methods used to obtain shock velocities and
  frequencies for pipe flow for the case of $D=20$~mm. (a) Modified
  density plot. (b-c) space-time correlation functions based on data
  shown in (a), for (b) $x=130$~cm and (c) $x=170$~cm. Note that the
  scales of (a) and (b,c) are different, so that the dominant angles
  appear different. (d) Thresholded data used for determination of
  shock frequency.}\label{f:pu}
\end{figure}

The method to obtain shock velocities for pipeflow data is illustrated
in Fig.~\ref{f:pu}.  First of all, to highlight the shocks that occur
in the density plots, some smoothing and differentiation is applied
yielding a field $s(x,t)$ as shown in Fig.~\ref{f:pu}(a). Clearly,
shocks are now visible as bright streaks (high values of $s$) in a
fairly even background. The temporal averages of the spatiotemporal
correlation function $C(x,\Delta x,\Delta t) :=\int dt ~s(x,t) s(x
+\Delta x,t+\Delta t)$, two examples of which are shown in
Fig.~\ref{f:pu}(b-c) for x=130 and 170 cm, clearly show a dominant
direction in space-time that can be associated with the local dominant
velocity of the shocks: this method of shock velocity determination
works over the whole range of parameters considered. In comparison to
the RGE method, this method is more local, and often gives a slightly
smaller estimate for the velocities (order of 10-20 \%). This is
presumably due to the fact that shocks have a tendency to perform
intermittent jumps forward, making their dominant local velocity
smaller than the typical velocity obtained over longer timescales.
Some effect of these jumps shows up in the correlation functions,
where the bright streak tends to bend for larger correlation distances
and time intervals.

To obtain the frequency from $s(x,t)$ is fairly straightforward: after
choosing an appropriate threshold, one obtains two-color images like
shown in Fig.~\ref{f:pu}(d). A simple algorithm suffices to count, for
a fixed value of $x$, the number of shock waves that occur.


\begin{thebibliography}{99}
\bibitem{Dnunote} This was not shown in~\cite{hd99}, but with improved data
analysis~\cite{RGEnote} of the old data we find a decent data collapse
when plotting $\nu(x)D$ vs. $w(x)/D$.
\bibitem{SSnote} The Shock Creation Rates shown here have been hand
counted, and every shock that propagates upwards at least 5 cm (no
matter its strength or continued existence) is counted. The shock
wave creation position statistics shown in Fig.~18 in \cite{hd99}
were based on a image processing method (DT - no longer used),
that only counted shocks of a certain strength. Consequently this
method showed a decreasing shock creation rate with increasing
$w(x)$ (for smooth walls).
\end{thebibliography}
\end{document}